\documentclass[11pt]{article}

\usepackage{amssymb}
\usepackage{amsmath}
\usepackage{amsthm}
\usepackage{geometry}
\theoremstyle{plain}

\newtheorem{lemma}{Lemma}

\begin{document}

\title{On the Convexity of $\log\det (I + K X^{-1})$}

\author{Young-Han Kim
and
Seung-Jean Kim%
\thanks{Email: yhk@ucsd.edu and sjkim@stanford.edu}}
\date{}
\maketitle

\begin{abstract}
A simple proof is given for the convexity of $\log \det (I+K X^{-1})$
in the positive definite matrix variable $X \succ 0$ with a given positive
semidefinite $K \succeq 0$.
\end{abstract}

%%{\it Index Terms}---Convexity, eigenvalue, log determinant, spectral
%%functions, symmetry.

\vspace{1em}

Convexity of functions of covariance matrices often plays an important
role in the analysis of Gaussian channels.  For example, suppose
$\mathbf{Y}$ and $\mathbf{Z}$ are independent complex Gaussian
$n$-vectors with $\mathbf{Y} \sim N(0, K)$ and $\mathbf{Z} \sim N(0,
X)$. Then,
\begin{equation}
\label{eq:func}
I(\mathbf{Y}; \mathbf{Y} + \mathbf{Z}) = \log \det (I + K X^{-1}).
\end{equation}
The following result is well known in the literature.
\begin{lemma}
\label{lem:main}
For a fixed $K \succeq 0$, $\log \det (I + K X^{-1})$ is convex in $X
\succ 0$, with strict convexity if $K \succ 0$.
\end{lemma}

A simple information theoretic proof was given by Diggavi and
Cover~\cite[Lemma II.3]{Diggavi--Cover2001} as a corollary to their
main saddle-point theorem stating that the worst additive noise is
Gaussian.  Very recently Mao \emph{et al.}~\cite{Mao--Su--Xu2006} gave a
different, but complicated proof, correcting an incomplete approach
taken in Kashyap \emph{et al.}~\cite{Kashyap--Basar--Srikant2004}.

The main purpose of this note is to introduce an elegant and simple
proof technique based on the theory of spectral functions of
Hermitian matrices, which will hopefully benefit other problems in
information theory with similar structure.

A real-valued function $f(X)$ of Hermitian argument $X \in
\mathbb{R}^{n\times n}$ is called a \emph{spectral function} if the
value of $f(X)$ depends only on (unordered) eigenvalues of $X$.  If
$\lambda(X) \in \mathbb{R}^n$ denotes the ordered eigenvalues of $X$
and $g$ is a real-valued symmetric (=permutation invariant) function
on $\mathbb{R}^n$, the composite function $(g\circ\lambda)(X) =
g(\lambda(X))$ is a spectral function.  Conversly, any spectral
function $f(X)$ can be decomposed in this way.  It is also easy to see
that $f(X)$ is a spectral function if and only if $f(X)$ is unitary
invariant, that is, $f(X) = f(Q X Q^\dagger)$ for any
unitary $Q$.

Convexity of a spectral function can be checked rather easily; a
spectral function $f = g\circ \lambda$ is (strictly) convex if and
only if the corresponding symmetric function $g$ is (strictly)
convex~\cite{Davis1957}.  In other wodrs, a spectral function $f(X)$
is convex for all Hermitian $X$ if and only if $f(X)$ is convex for
all real diagonal $X$.  Examples of convex spectral functions include
the trace, the largest eigenvalue, and the sum of the $k$ largest
eigenvalues, of a Hermitian matrix; and the trace of the inverse of a
positive definite matrix (as well as the log determinant of the
inverse as will be shown shortly).  For a few other interesting
properties of spectral functions, refer to \cite[Section
5.2]{Borwein--Lewis2006}.

Now we are ready for a ``two-line'' proof.

\begin{proof}[\bf Proof of Lemma~\ref{lem:main}]
Since $X \mapsto K^{1/2}XK^{1/2}$ is a positive linear map (strict if
$K \succ 0$) and
\[
\log \det (I + K X^{-1}) = \log \det(I + K^{1/2} X^{-1} K^{1/2}),
\]
it suffices to establish the strict convexity of
\[
f(X) = \log \det (I + X^{-1})
\]
in $X$.  (See \cite[Section 3.2.4]{Boyd--Vandenberghe2004} for the
composition rule for convexity.)

Now it is easy to check that $f(X)$ is a {spectral} function.
Indeed,
\[
f(X) = g(\lambda(X)) := \sum_{i=1}^n \log \left ( 1 +
\frac{1}{\lambda_i(X)} \right),
\]
where $\lambda_1(X), \ldots, \lambda_n(X) > 0$ are the eigenvalues of
$X$.  But trivially $\log (1 + 1/t)$ is strictly convex for $t > 0$,
which implies the strict convexity of $g$ and thus of $f$.
\end{proof}

\def\cprime{$'$} \def\cprime{$'$} \def\cprime{$'$} \def\cprime{$'$}


\begin{thebibliography}{1}
\providecommand{\url}[1]{#1}
\csname url@rmstyle\endcsname
\providecommand{\newblock}{\relax}
\providecommand{\bibinfo}[2]{#2}
\providecommand\BIBentrySTDinterwordspacing{\spaceskip=0pt\relax}
\providecommand\BIBentryALTinterwordstretchfactor{4}
\providecommand\BIBentryALTinterwordspacing{\spaceskip=\fontdimen2\font plus
\BIBentryALTinterwordstretchfactor\fontdimen3\font minus
  \fontdimen4\font\relax}
\providecommand\BIBforeignlanguage[2]{{%
\expandafter\ifx\csname l@#1\endcsname\relax
\typeout{** WARNING: IEEEtran.bst: No hyphenation pattern has been}%
\typeout{** loaded for the language `#1'. Using the pattern for}%
\typeout{** the default language instead.}%
\else
\language=\csname l@#1\endcsname
\fi
#2}}

\bibitem{Borwein--Lewis2006}
J.~M. Borwein and A.~S. Lewis, \emph{Convex Analysis and Nonlinear
  Optimization: Theory and Examples}, 2nd~ed.\hskip 1em plus 0.5em minus
  0.4em\relax New York: Springer, 2006.

\bibitem{Boyd--Vandenberghe2004}
S.~Boyd and L.~Vandenberghe, \emph{Convex Optimization}.\hskip 1em plus 0.5em
  minus 0.4em\relax Cambridge: Cambridge University Press, 2004.

\bibitem{Davis1957}
C.~Davis, ``All convex invariant functions of hermitian matrices,'' \emph{Arch.
  Math.}, vol.~8, pp. 276--278, 1957.

\bibitem{Diggavi--Cover2001}
S.~N. Diggavi and T.~M. Cover, ``The worst additive noise under a covariance
  constraint,'' \emph{{IEEE} Trans. Inform. Theory}, vol. IT-47, no.~7, pp.
  3072--3081, Nov. 2001.

\bibitem{Kashyap--Basar--Srikant2004}
A.~Kashyap, T.~Ba{\c{s}}ar, and R.~Srikant, ``Correlated jamming on {MIMO}
  {G}aussian fading channels,'' \emph{{IEEE} Trans. Inform. Theory}, vol.
  IT-50, no.~9, pp. 2119--2123, Sept. 2004.

\bibitem{Mao--Su--Xu2006}
W.~Mao, X.~Su, and X.~Xu, ``Comments on ``{C}orrelated jamming on {MIMO}
  {G}aussian fading channels'','' \emph{{IEEE} Trans. Inform. Theory}, vol.
  IT-52, no.~11, pp. 5163--5165, Nov. 2006.

\end{thebibliography}
\end{document}